\documentclass[twocolumn,english,aps,preprintnumbers,amsmath,amssymb,superscriptaddress]{revtex4-2}
\usepackage[latin9]{inputenc}
\usepackage{amsmath}
\usepackage{graphicx}

\makeatletter
\@ifundefined{textcolor}{}
{%
 \definecolor{BLACK}{gray}{0}
 \definecolor{WHITE}{gray}{1}
 \definecolor{RED}{rgb}{1,0,0}
 \definecolor{GREEN}{rgb}{0,1,0}
 \definecolor{BLUE}{rgb}{0,0,1}
 \definecolor{CYAN}{cmyk}{1,0,0,0}
 \definecolor{MAGENTA}{cmyk}{0,1,0,0}
 \definecolor{YELLOW}{cmyk}{0,0,1,0}
}

\def\lsco{La$_{2-x}$Sr$_x$CuO$_4$}

\usepackage{dcolumn}
\usepackage{bm}
\usepackage{color}

\makeatother

\usepackage{babel}

\begin{document}

\preprint{preprint(\today)}

\title{Designing the stripe-ordered cuprate phase diagram through uniaxial-stress}

\author{Z.~Guguchia$^{\dag}$}
\email{zurab.guguchia@psi.ch} 
\affiliation{Laboratory for Muon Spin Spectroscopy, Paul Scherrer Institute, CH-5232
Villigen PSI, Switzerland}

\author{D.~Das}
\thanks{These authors contributed equally.}
\affiliation{Laboratory for Muon Spin Spectroscopy, Paul Scherrer Institute, CH-5232
Villigen PSI, Switzerland}

\author{G.~Simutis}
\thanks{These authors contributed equally.}
\affiliation{Laboratory for Neutron and Muon Instrumentation, Paul Scherrer Institut, CH-5232 Villigen PSI, Switzerland}

\author{T.~Adachi}
\affiliation{Department of Engineering and Applied Sciences, Sophia University, 7-1 Kioi-cho, Chiyoda-ku, Tokyo 102-8554, Japan}

\author{J. K\"{u}spert}
\affiliation{Physik-Institut, Universit\"{a}t Z\"{u}rich, Winterthurerstrasse 190, CH-8057 Z\"{u}rich, Switzerland}%

\author{N. Kitajima}
\affiliation{Department of Applied Physics, Tohoku University, 6-6-05 Aoba, Aramaki, Aoba-ku, Sendai 980-8579, Japan}

\author{M.~Elender}
\affiliation{Laboratory for Muon Spin Spectroscopy, Paul Scherrer Institute, CH-5232 Villigen PSI, Switzerland}

\author{V. Grinenko}
\affiliation{Tsung-Dao Lee Institute, Shanghai Jiao Tong University, Pudong, 201210 Shanghai, China}

\author{O.~Ivashko}
\affiliation{Deutsches Elektronen-Synchrotron DESY, Notkestraße 85, 22607 Hamburg, Germany}

\author{M.v.~Zimmermann}
\affiliation{Deutsches Elektronen-Synchrotron DESY, Notkestraße 85, 22607 Hamburg, Germany}

\author{M. M\"{u}ller}
\affiliation{Paul Scherrer Institute, Condensed Matter Theory, PSI Villigen, Switzerland}

\author{C.~Mielke III}
\affiliation{Laboratory for Muon Spin Spectroscopy, Paul Scherrer Institute, CH-5232 Villigen PSI, Switzerland}

\author{F.~Hotz}
\affiliation{Laboratory for Muon Spin Spectroscopy, Paul Scherrer Institute, CH-5232 Villigen PSI, Switzerland}

\author{C. Mudry}
\affiliation{Paul Scherrer Institute, Condensed Matter Theory, PSI Villigen, Switzerland}
\affiliation{Institut de Physique, EPF Lausanne, Lausanne, CH-1015, Switzerland}

\author{C.~Baines}
\affiliation{Laboratory for Muon Spin Spectroscopy, Paul Scherrer Institute, CH-5232 Villigen PSI, Switzerland}

\author{M.~Bartkowiak}
\affiliation{Laboratory for Neutron and Muon Instrumentation, Paul Scherrer Institut, CH-5232 Villigen PSI, Switzerland}

\author{T.~Shiroka}
\affiliation{Laboratory for Muon Spin Spectroscopy, Paul Scherrer Institute, CH-5232 Villigen PSI, Switzerland}

\author{Y. Koike}
\affiliation{Department of Applied Physics, Tohoku University, 6-6-05 Aoba, Aramaki, Aoba-ku, Sendai 980-8579, Japan}

\author{A.~Amato}
\affiliation{Laboratory for Muon Spin Spectroscopy, Paul Scherrer Institute, CH-5232 Villigen PSI, Switzerland}

\author{C.W.~Hicks}
\affiliation{Max Planck Institute for Chemical Physics of Solids, D-01187 Dresden, Germany}

\author{G.D.~Gu}
\affiliation{Condensed Matter Physics and Materials Science Division, Brookhaven National Laboratory, Upton, NY 11973, USA}

\author{J.M.~Tranquada}
\affiliation{Condensed Matter Physics and Materials Science Division, Brookhaven National Laboratory, Upton, NY 11973, USA}

\author{H.-H. Klauss}
\affiliation{Institute for Solid State and Materials Physics, Technische Universitat Dresden, D-01069 Dresden, Germany}

\author{J.J. Chang}
\affiliation{Physik-Institut, Universit\"{a}t Z\"{u}rich, Winterthurerstrasse 190, CH-8057 Z\"{u}rich, Switzerland}%

\author{M.~Janoschek}
\affiliation{Laboratory for Neutron and Muon Instrumentation, Paul Scherrer Institut, CH-5232 Villigen PSI, Switzerland}%

\author{H.~Luetkens}
\email{hubertus.luetkens@psi.ch} 
\affiliation{Laboratory for Muon Spin Spectroscopy, Paul Scherrer Institute, CH-5232 Villigen PSI, Switzerland}

\begin{abstract}

\bf{The ability to efficiently control charge and spin in the cuprate high-temperature superconductors 
is crucial for fundamental research and underpins technological development. Here, we explore the tunability of magnetism, superconductivity and crystal structure in the stripe phase of the cuprate La$_{2-x}$Ba$_{x}$CuO$_{4}$, with $x$ = 0.115 and 0.135, by employing temperature-dependent (down to 400 mK) muon-spin rotation and AC susceptibility, as well as X-ray scattering experiments under compressive uniaxial stress in the CuO$_{2}$ plane. A sixfold increase of the 3-dimensional (3D) superconducting critical temperature $T_{\rm c}$ and a full recovery of the 3D phase coherence is observed in both samples with the application of extremely low uniaxial stress of ${\sim}$ 0.1 GPa. 
This finding demonstrates the removal of the well-known 1/8-anomaly of cuprates by uniaxial stress. On the other hand, the spin-stripe order temperature as well as the magnetic fraction at 400 mK show only a modest decrease under stress. Moreover, the onset temperatures of 3D superconductivity and spin-stripe order are very similar in the large stress regime. However, a substantial decrease of the magnetic volume fraction and a full suppression of the low-temperature tetragonal structure is found at elevated temperatures, which is a necessary condition for the development of the 3D superconducting phase with optimal $T_{\rm c}$. Our results evidence a remarkable cooperation between the long-range static spin-stripe order and the underlying crystalline order with the three-dimensional fully coherent superconductivity. Overall, these results suggest that the stripe- and the SC order may have a common physical mechanism.}

\end{abstract}


\maketitle

\section{Introduction}

\begin{figure*}[t]
\centering
\includegraphics[width=1.0\linewidth]{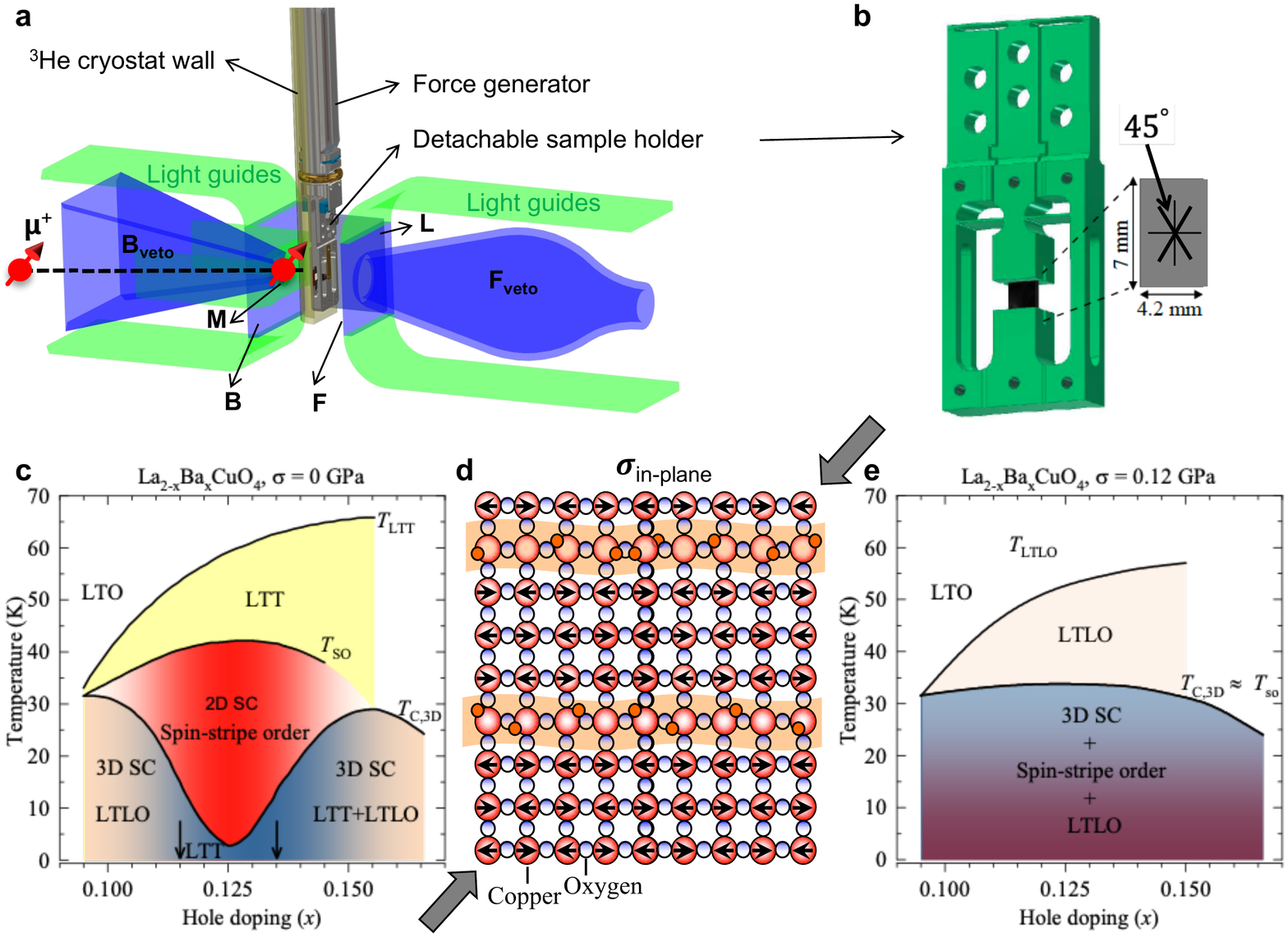}
\vspace{-0.5cm}
\caption{\textbf{A schematic overview of the experimental setup and phase diagram.} 
(a) A schematic overview of the experimental setup and the uniaxial stress cell. Spin polarized muons with spin $S_{\mu}$ at $45^{\circ}$ with respect to the $c$-axis of the crystal, are implanted in the sample. The sample is surrounded by a backward veto detector B$_{veto}$, 
a cup-shaped forward veto detector F$_{veto}$ and four positron detectors: Forward (F), Backward (B), Left (L), and Right (R - not shown for clarity). An electronic clock starts when a muon passes through the muon detector (M) and stops once a decay positron is detected in the positron detectors. B$_{veto}$ consists of a hollow scintillator pyramid with a 7${\times}$7 mm hole facing the M counter. The purpose of B$_{veto}$ is to collimate the muon beam to a 7${\times}$7 mm spot and to reject muons (and their decay positrons) missing the aperture.
F$_{veto}$ rejects the muons (and their decay positrons) that miss the sample. (b) The uniaxial stress sample holder, used for the ${\mu}$SR experiments. (c) The schematic temperature-doping phase diagram of  La$_{2-x}$Ba$_{x}$CuO$_{4}$ for zero-stress ${\sigma}$ = 0 GPa. The arrows indicate the present doping values. The various phases in the phase diagram are denoted as follows: Low-temperature orthorhombic (LTO), low-temperature tetragonal (LTT), low-temperature less orthorhombic (LTLO), spin-stripe order (SO), 2D superconductivity (2D SC) and 3D superconductivity (3D SC). (d) Illustration of a domain of spin- and charge-stripe order for a layer of LBCO, indicating the periods of the charge (4$\it{a}$) and spin (8$\it{a}$) modulations. The compressive stress was applied at an angle of 45$^\circ$ to the Cu-O bond direction, as indicated by the gray arrows. (e) The hypothetical temperature-doping phase diagram of  La$_{2-x}$Ba$_{x}$CuO$_{4}$, expected under applied stress of ${\sigma}$ = 0.12 GPa in the region around $x$ = 1/8 doping. This phase diagram implies that the stress enhances  the 3D SC critical temperature $T_{\rm c,3D}$ and reduces slightly the spin-stripe order temperature $T_{\rm so}$ until they acquire similar values $T_{\rm c,3D}$ ${\simeq}$ $T_{\rm so}$. Stress also causes the reduction of the spin-stripe-ordered volume fraction at elevated temperatures, and the full suppression of the LTT phase.} 
\label{fig1}
\end{figure*}

High-transition-temperature (high-$T_{\rm c}$) superconductivity in copper oxides (cuprates) \cite{Bednorz,Robinson,Agterberg,Berg1,QLi,Abanov,Vekhter,Dahm} is one of the most intriguing emergent phenomena in strongly-correlated electron systems. Besides superconductivity, the phase diagram of some cuprates includes spin- and charge order in patterns of alternating stripes \cite{Tranquada1,Huecker,Tranquada2,Abbamonte,Vojta,Kivelson}. There is an increasingly strong evidence of static/fluctuating stripe correlations in superconducting cuprates  \cite{Julien,Chang,Tranquadareview}. One of the most astonishing manifestations of the competition between the various ground states in cuprates occurs in the prototypical cuprate superconductor La$_{2-x}$Ba$_{x}$CuO$_{4}$ \cite{Tranquada1,Huecker,Tranquada2}. It exhibits an anomalous suppression of the uniform 3D bulk superconductivity when the hole concentration $x$ is near 1/8, where static charge- and spin-stripe orders and a structural phase transition [from a low temperature orthorhombic (LTO) to a low-temperature-tetragonal (LTT) phase] occur simultaneously. \cite{Axe,Klauss,KlaussPRL}. In the LTT phase, the CuO$_{6}$ octahedra rotate about alternate orthogonal axes ([100] and [010]) in successive layers. The atomic displacements in the LTO structure form a diagonal pattern, whereas, in the LTT case, the pattern of displacements is horizontal (or vertical). Thus, it is believed that, in the LTT phase, the horizontal stripes are pinned by the lattice potential, which is responsible for the orthogonal stripe order (alternate switching of the stripe direction from plane to plane) along the $c$ axis. While 3D superconductivity is strongly suppressed near $x$ = 1/8, 2D superconductivity has the same onset temperature as spin-stripe order \cite{Tranquada2008,Li,GuguchiaPRB}. To explain 2D superconductivity, a pair-density-wave (PDW) order \cite{Fradkin,Himeda} (i.e., spatially modulated SC order) between the charge stripes has been proposed. In this state, the superconducting wave function oscillates from positive to negative from one charge stripe to the next. Because the pinning of the charge stripes by the lattice anisotropy rotates 90$^\circ$ from one layer to the next, the interlayer Josephson coupling is frustrated. While this frustration inhibits the development of 3D superconducting order, it has no restriction on 2D order. Thus, PDW order is considered to be compatible with both the charge and spin-stripe orders. It is believed that uniform $d$-wave and striped PDW orders are competing SC orders \cite{Robinson,Agterberg}. One of the long-standing mysteries in cuprates is whether the stripe- and the superconducting states involve distinct electron-pairing mechanisms and what mechanisms control the competition between such states.

\begin{figure*}[t!]
\centering
\includegraphics[width=0.9\linewidth]{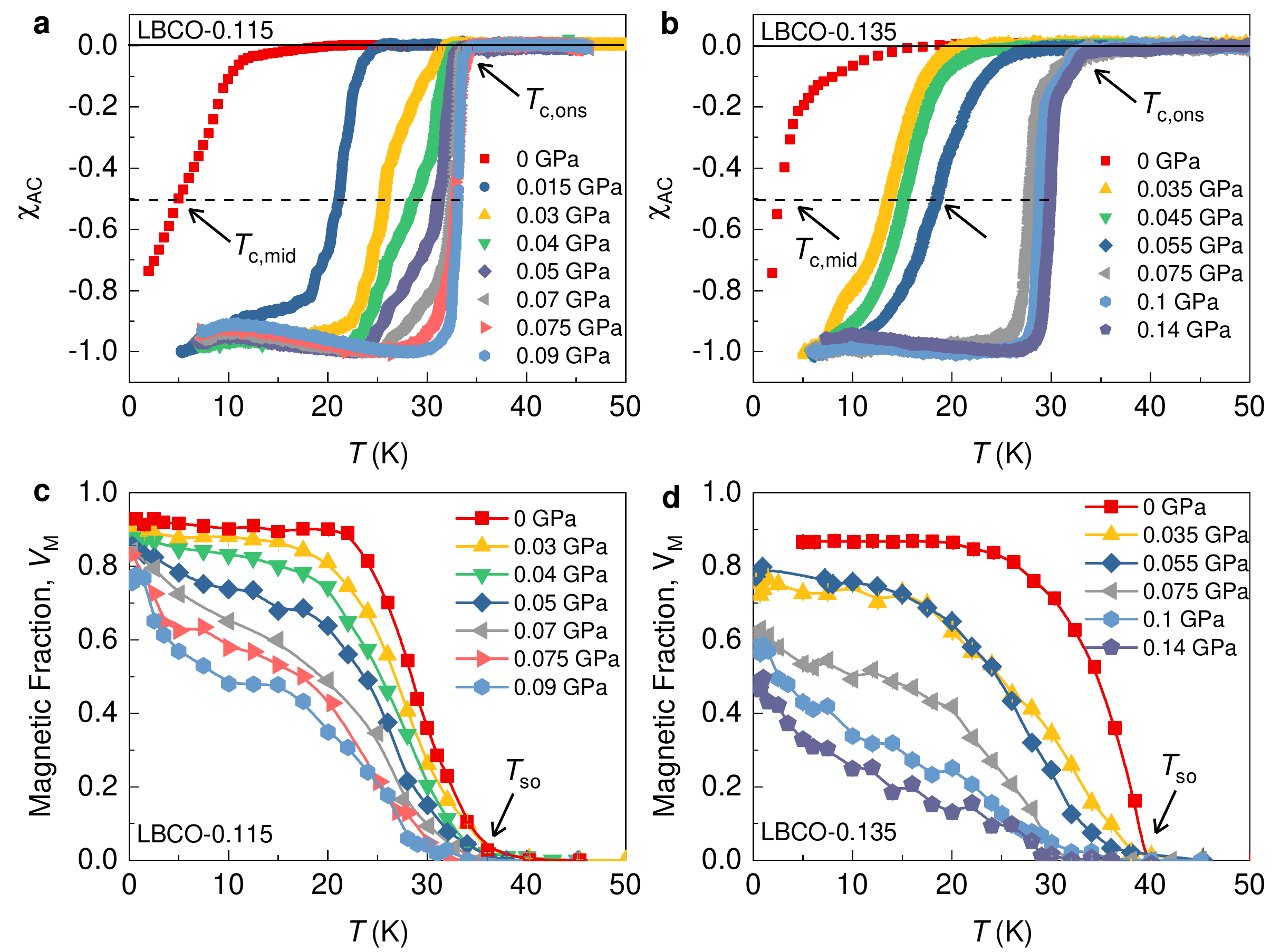}
\vspace{-0.3cm}
\caption{\textbf{Superconducting screening and ordered magnetic fraction for LBCO under stress.} 
(a-b) The temperature dependence of the (dia)magnetic susceptibility for LBCO $x=0.115$ (a) and LBCO $x=0.135$ (b), recorded at ambient and under various degrees of compressive stress. Arrows mark the onset temperature $T_{\rm c,ons}$ and the temperature $T_{\rm c,mid}$ at which ${\chi'}_{\rm dc} = -0.5$. (c-d) The temperature dependence of the magnetically ordered volume fraction for LBCO $x=0.115$ (c) and LBCO $x=0.135$ (d) recorded under various stress values.} 
\label{fig1}
\end{figure*}

Hydrostatic pressure has long been considered as a way of tuning the stripe phase in cuprates, but it was found that the pressure effect on the 3D superconducting transition temperature in LBCO near 1/8 is quite modest, even with the loss of the long-range LTT structure \cite{HuckerPRL,GuguchiaNPJ}. In-plane uniaxial stress has recently been shown to be a more efficient control parameter \cite{GuguchiaPRLstrain,Fujita2004,Boyle,SimutisXray,SimutisMazzone,QWang,KimYBCO,JChoi,Kamminga}. It was shown that the 2D superconductivity in LBCO ($x$ = 0.115) can be pushed toward 3D order by the application of strain \cite{GuguchiaPRLstrain}. 
Strain was also shown to affect spin-stripe order in LBCO ($x$ = 0.115). However, the generic character of such observations in a broader region around 1/8 doping is not without ambiguity.

Here, we use in-plane compressive uniaxial stress applied to the CuO$_2$ layers to perturb the stripe phase of the LBCO phase diagram. An {\it in situ} piezoelectrically-driven stress device was used to microscopically probe the spin-stripe order with muon spin rotation ($\mu$SR) spectroscopy (Figs. 1a and b), and the superconducting transitions with ac susceptibility in single crystalline samples of  La$_{2-x}$Ba$_{x}$CuO$_{4}$ with $x$ = 0.115 (below 1/8 doping) and $x$ = 0.135 (above 1/8 doping). The details of the $\mu$SR technique, data analysis, and the uniaxial stress devices are published elsewhere \cite{GuguchiaPRLstrain}. The stress effect on the LTO-to-LTT transition in the sample $x$ = 0.115 is probed by X-ray scattering using a recently designed in-situ uniaxial pressure device \cite{SimutisXray}. The details of the sample mounting are given in the methods section. We show that an extremely low in-plane uniaxial stress of ${\sim}$ 0.1 GPa substantially modifies the temperature-doping phase diagram of LBCO around 1/8-doping (see Figure 1c-e), leading to a phase diagram with no long-range LTT phase (assumed necessary to pin the  stripes) and no dip in the 3D $T_{\rm c}$, while still preserving the spin-stripe order with $T_{\rm so}$ ${\simeq}$ $T_{\rm c,3D}$. Since the stress required to establish the phase diagram shown in Figure 1c is very small, we conclude that the spin-stripe order, the spatially-modulated 2D (PDW) and uniform 3D SC orders are energetically very finely balanced and have a similar pairing mechanism.

\section{Results}
\subsection{Probing superconductivity under stress}

To monitor the effect of stress on superconductivity in both LBCO-0.115 and LBCO-0.135 (applied at an angle of 45$^\circ$ to the Cu-O bond direction), {\it in situ} ac susceptibility measurements were performed. An excitation magnetic field was applied approximately along the $c$ axis, either just before or after the $\mu$SR measurements, at each stress value. The results are shown in Figs.~2a and b. The diamagnetic response of both crystals at 0 GPa corresponds to the measurements before mounting them in the stress apparatus. The samples at zero-pressure were zero-field cooled and then measured in a dc field of $\mu_0 H = 0.5$~mT. The field was applied parallel to the CuO$_2$ planes, so that the resulting shielding currents had to flow between the layers, making the measurement sensitive to the onset of 3D superconductivity below ${\sim}$ 11 K for LBCO-0.115 and below ${\sim}$ 7 K for LBCO-0.135, consistent with previous work \cite{Tranquada2008,GuguchiaPRL2017,AdachiPRB2001}. The onset of weak diamagnetism near 22 K corresponds to the 2D superconducting order, as it was previously discussed. To characterize the changes in the superconducting critical temperature, we identify the onset temperature $T_{c,{\rm ons}}$ (which equals $T_{\rm c,2D}$ at zero stress) and the midpoint temperature $T_{c,{\rm mid}}$ (which is a good measure of 3D SC order temperature $T_{\rm c,3D}$), as indicated in Figs.~2a and b, and with the strongest diamagnetic response with 100 ${\%}$-volume-fraction superconductivity. Remarkably, the compressive stress causes a rapid rise of $T_{c,{\rm mid}}$ from 5 to 32~K in LBCO-0.115 and from 3 to 30~K in LBCO-0.135, where $T_{c,{\rm mid}}$ saturates. The change in $T_{c,{\rm ons}}$ is much more modest. Namely, $T_{c,{\rm ons}}$ increases from 20 to 32~K. Consequently, the bulk transition $T_{\rm c,3D}$ of LBCO around 1/8-doping  rises from a very suppressed value to the one that is quite similar to the optimal value of SC critical temperature observed in LBCO or \lsco\ (LSCO) at the same doping level \cite{Takagi1989}.

\subsection{Probing the spin-stripe order under stress}

A combination of weak transverse-field (TF) and zero-field (ZF) $\mu$SR measurements were carried out to characterise the evolution of the spin-stripe order with compressive stress using the same device. In a weak-TF measurement, muons in regions without a local magnetic order precess in the small applied field. Muons that stop in regions with magnetic order and, therefore, experience the vector sum of external and internal fields, dephase rapidly. This causes a rapid reduction in the observable polarization and allows us to determine the magnetically ordered volume fraction $V_{\rm M}$. The temperature dependence of $V_{\rm M}$ for various stress values is presented in Figs.~2c and d, for LBCO-0.115 and LBCO-0.135 samples, respectively. Upon increasing stress there is a decrease in the spin-ordering temperature $T_{\rm so}$, from ${\sim}$ 38 K at 0 GPa to ${\sim}$ 30 K at 0.09 GPa in both samples. The magnetic volume fraction $V_{\rm M}$ decreases much more steeply and the effect is stronger in LBCO-0.135 as compared to LBCO-0.115. At 10 K, $V_{\rm M}$ decreases by factor of two in LBCO-0.115 and by a factor of three in LBCO-0.135. However, the magnetic fraction shows a clear upturn below 10 K, as one can see in Figs. 2c and d. In LBCO-0.115 at 0.4 K, $V_{\rm M}$ reaches 80 ${\%}$ even at the highest applied stress of 0.09 GPa, which is nearly the same as the one at ambient conditions. These results show that while $V_{\rm M}$ is strongly suppressed at elevated temperatures, it tends to recover below 10 K upon approaching the zero-temperature. Therefore, nearly 100 ${\%}$ superconductivity coexists with nearly 100 ${\%}$ spin-stripe order in the zero-temperature (i.e., quantum) limit.

\begin{figure*}[t!]
\includegraphics[width=1.0\linewidth]{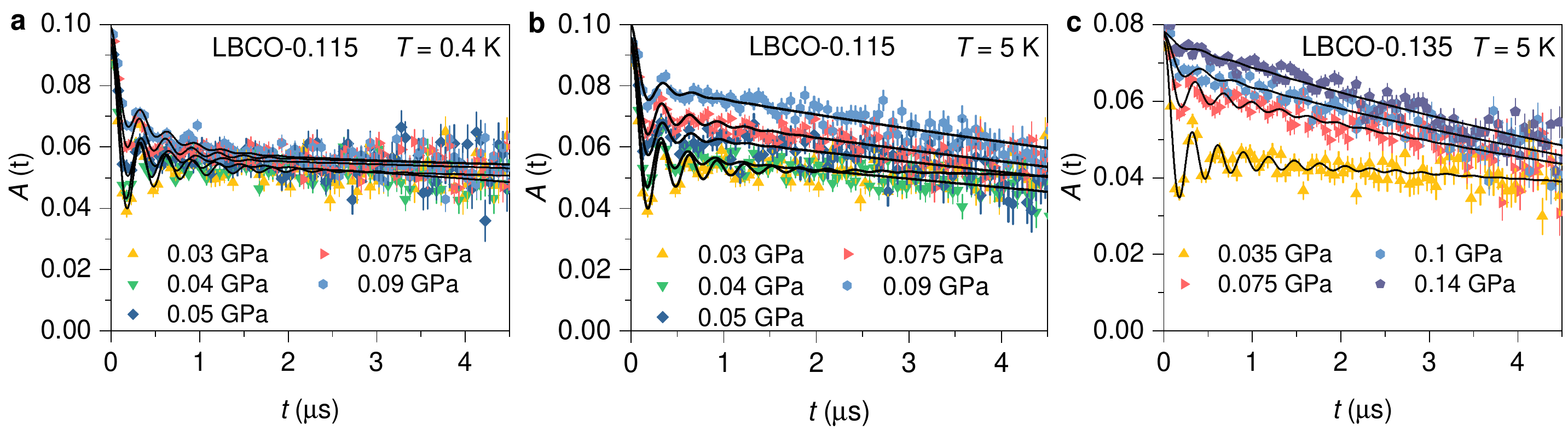}
\vspace{-0.5cm}
\caption{ \textbf{${\mu}$SR time spectra for LBCO.} The zero-field ${\mu}$SR spectra for LBCO-0.115, recorded at the base temperature $T$ = 0.4 K (a) and at $T$ = 5 K (b) under various stress values. (c) The zero-field ${\mu}$SR spectra for LBCO-0.135 recorded at $T$ = 5 K under various stress values.}   
\label{fig1}
\end{figure*}

In ZF $\mu$SR measurements, the muon spins precess exclusively in the internal local field associated with the static long-range magnetic order. As shown in Fig.~3a-c, several oscillations remain clearly observable under increasing compressive stress values, despite the reduction in magnetic volume fraction (amplitude of the signal). This implies that the spin-stripe order remains long-range even at the highest applied stress. The oscillation frequency and, thus, the characteristic internal field $B_{\rm int}$ at the muon stopping site, is only weakly affected by stress.

\begin{figure*}[t!]
\centering
\includegraphics[width=0.8\linewidth]{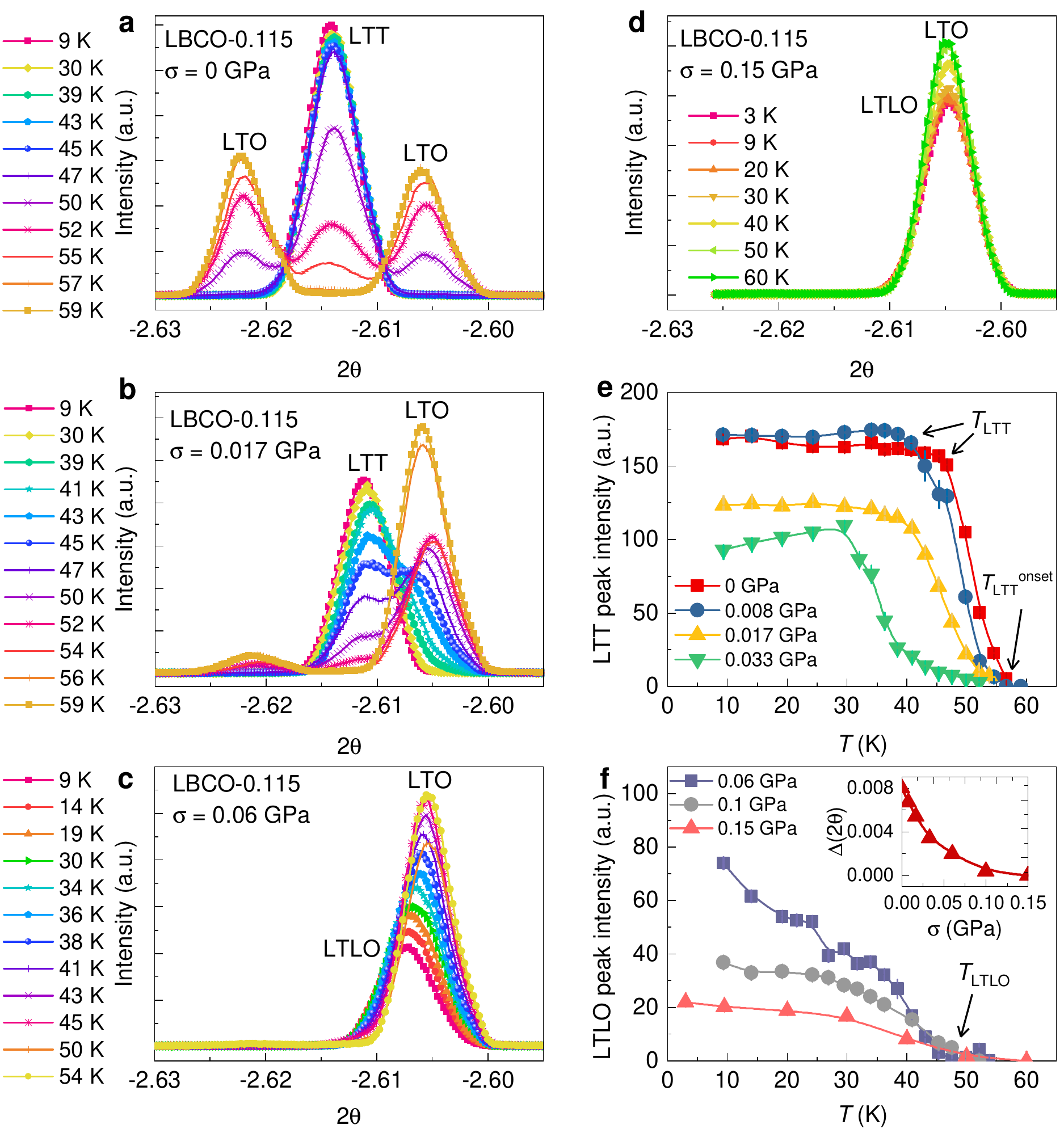}
\vspace{-0.2cm}
\caption{(a) \textbf{Structural transition in LBCO under uniaxial stress.} 
Scans of the (020)$_{o}$/(200)$_{o}$ Bragg reflections at several temperatures through the transition from the LTO phase to the LTT phase, measured under an uniaxial stress of 0 GPa (a), 0.017 GPa (b), 0.06 GPa (c), and 0.15 GPa (d). The temperature dependences of the LTT (e) and LTLO (f) peak intensities for LBCO-0.115, measured under various stress values. The inset shows the difference between the scattering angle measured at $T$ ${\textgreater}$ 50 K and that at base-$T$ of 9 K, i.e., ${\Delta}$(2${\theta}$) = ${2\theta}$($T$ ${\textgreater}$ 50 K) - ${2\theta}$($T$ = 9 K).}
\label{fig1}
\end{figure*}

\subsection{Probing the crystal structure under stress}

$\mu$SR and AC susceptibility experiments presented above allow us to study the stress effect on superconductivity and spin-stripe order. 
In order to interpret the results, it is also necessary to characterize the underlying crystalline order \cite{Frison}. To study the uniaxial stress evolution of the structural phase transition from a LTO to a low temperature tetragonal (LTT) phase, we performed X-ray scattering experiments at the P21.1 beamline at PETRA-III at DESY in Hamburg, Germany. The crystal was positioned with tetragonal [0,0,1] and [1,1,0] directions spanning the scattering plane and the stress was applied at an angle of 45$^\circ$ to the Cu-O bond direction. Fig. 4a shows the high-resolution elastic scattering scans through the (020)$_{o}$/(200)$_{o}$ (refers to orthogonal notation) Bragg peaks at ambient pressure for various temperatures for LBCO-0.115. At 59 K, the LTO peaks are sharp, well separated and have equal intensities, implying equal volume fractions for the respective orthorhombic domains. The LTT peak begins to appear below 57 K. Then, there is a coexistence of the LTO and LTT peaks down to approximately 47 K and below this temperature a single strong LTT peak is observed. The scans for selected applied stresses are shown in Figures 4b-d. As one can see in Fig. 4b, when a stress of 0.017 GPa is applied the intensity of the left LTO peak is suppressed, indicating the structural detwinning of the orthorhombic domains. At this stress value, the appearance of LTT peak is also seen below 54 K. We note that the stress also modifies the scattering angle for the LTT peak, such that it shifts towards the LTO peak. In Figure 4e, we plot the temperature dependence of the integrated intensity of the LTT peak, measured for various uniaxial stresses. To characterize the changes in the structural transition temperatures, we identify the onset temperature $T_{{\rm LTT, ons}}$ (below which the LTT phase starts to appear and coexists with the LTO phase) and the temperature $T_{\rm LTT}$ (below which only LTT phase is observed), as indicated in Fig. 4e. This plot indicates that at ambient pressure the onset of the LTO-to-LTT transition is $T_{{\rm LTT, ons}}$ ${\simeq}$ 57 K, where the two phases coexist within a 10 K temperature range. For 0.017 GPa, $T_{{\rm LTT, ons}}$ is slightly reduced and there is a broader temperature region of coexistence. When stress is increased to 0.033 GPa, the onset of the LTT-LTO transition remains nearly unchanged, the coexistence region is further broadened and the LTT peak shifts further towards the LTO peak. As one can see in Figure 4c, upon increasing the stress to 0.06 GPa, a sharp LTO peak is observed from 54 K down to 45 K. Then, upon further lowering the temperature, there is no development of an LTT peak. Instead, the intensity of the LTO peak decreases and it shifts slightly towards higher scattering angles. We interpret this behaviour as the occurrence of an intervening LTLO phase and the intensity variation can be understood as a reduction in orthorhombicity. Thus, the scattering intensity seems to provide a precise measure of the LTO-LTLO structural change. The temperature dependences of the LTLO intensities (integrated intensity after subtracting the intensity at $T$ ${\textgreater}$ 50 K, where system is in the LTO phase) for 0.06 GPa and higher stresses are shown in Figure 4f. Here, the inset shows the difference between the scattering angle measured at $T$ ${\textgreater}$ 50 K and that at base-$T$ of 9 K, i.e., ${\Delta}$(2${\theta}$) = ${2\theta}$($T$ ${\textgreater}$ 50 K) - ${2\theta}$($T$ = 9 K). When applying stresses higher than 0.06 GPa the LTO peak intensity and the scattering angle are less affected and show only a weak temperature dependence, as shown in Fig. 4d and f. These results suggest that, at low temperatures, the stress tends to stabilise the full volume LTLO phase.

\begin{figure*}[t!]
\centering
\includegraphics[width=1.0\linewidth]{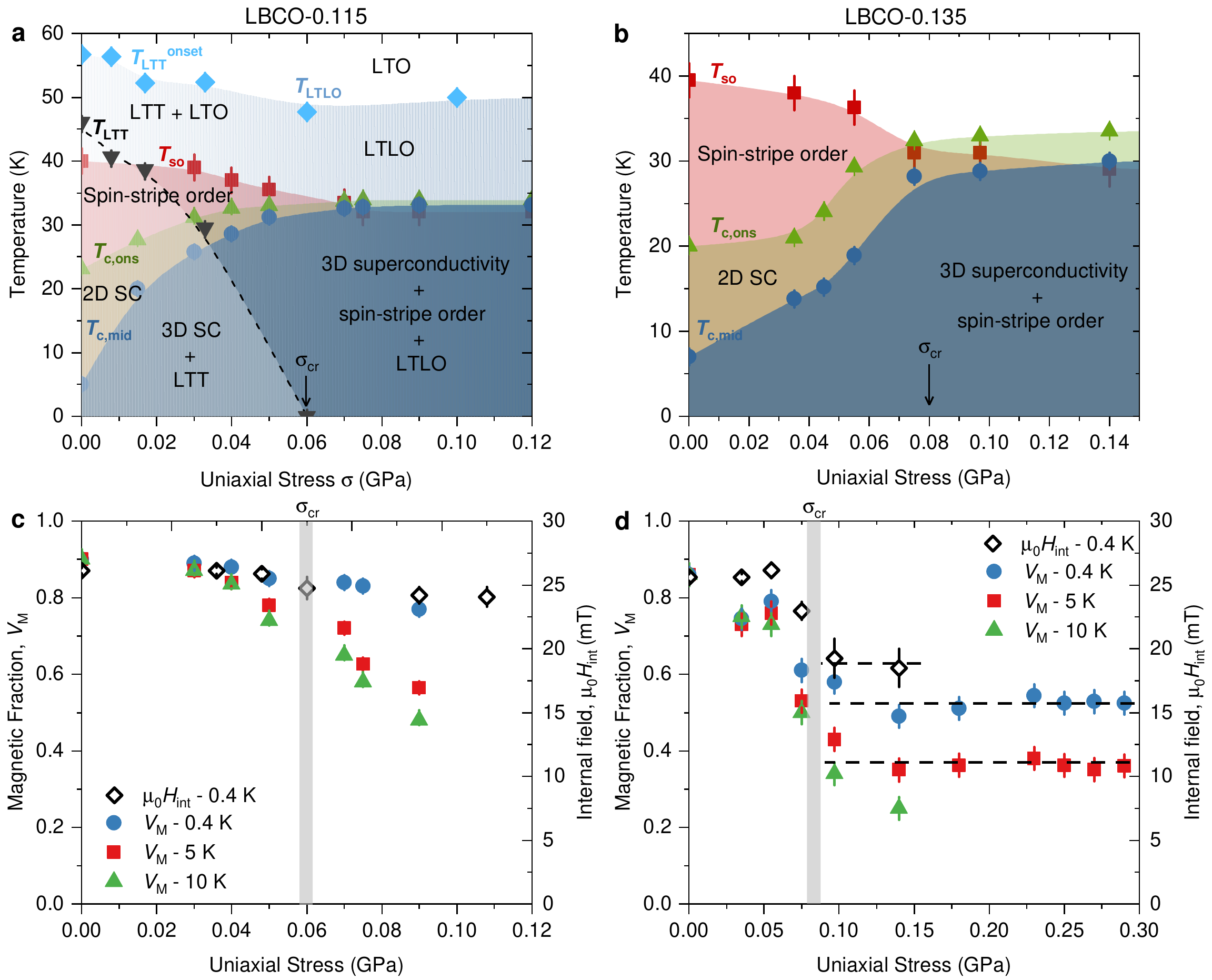}
\vspace{-0.2cm}
\caption{(a) \textbf{Phase diagrams of LBCO under uniaxial stress.} 
Dependence of the SC transition temperatures ($T_{c,2D}$, $T_{c,3D}$) and of static spin-stripe order temperature ($T_{\rm so}$) on compressive stress in LBCO $x = 0.115$ (a) and in LBCO $x = 0.135$ (b). The stress dependence of the structural transition temperatures ($T_{\rm LTT}$, $T_{\rm LTO}$) for  
LBCO $x = 0.115$ is also shown in panel (a). Black arrow marks the critical stress value ${\sigma}_{cr}$, above which a sharp 3D SC transition is established.
(c-d) The stress dependence of the internal magnetic field $B_{\rm int}$ for the base-$T$ of 0.4 K and of magnetically ordered fraction $V_{\rm M}$ for various temperatures $T$ = 0.4 K, 5 K, and 10 K for LBCO $x = 0.115$ (c) and LBCO $x = 0.135$ (d). Vertical grey lines mark the critical stress value ${\sigma}_{cr}$.}
\label{fig1}
\end{figure*}

\subsection{Phase Diagram}

Our overall results are summarized in Fig.~5. The spin-stripe order temperature $T_{\rm so}$, the LTO-to-LTT/LTLO structural phase transition temperature and superconducting transition temperatures are plotted against stress in Fig.~5a and b for LBCO-0.115 and LBCO-0.135, respectively. The stress dependence of the low-temperature value of magnetic volume fraction and internal magnetic field at 0.4 K are shown in Fig.~5c and d. Figure 5a and b show that the crossover from 2D to 3D superconducting order occurs at a characteristic critical uniaxial stress of ${\sigma}_{cr}$ = 0.06 GPa for LBCO-0.115 and 0.08 GPa for LBCO-0.135. Remarkably, establishment of optimal 3D SC order is followed by the full suppression of the LTT structure at ${\sigma}_{cr}$ = 0.06 GPa, which is replaced by the LTLO structure, as it was demonstrated for the $x$=0.115 sample. Interestingly, $T_{\rm so}$ decreases by small amount to essentially match $T_{c,{\rm mid}}$ for ${\sigma}$ ${\textgreater}$ ${\sigma}_{cr}$ for both samples. We note that, only a modest stress-induced decrease in $B_{\rm int}$ (Fig. 5c and d) is resolved, indicating that the magnetic structure is well ordered also under stress. The dominant change of the spin-stripe order induced by uniaxial stress is a strong reduction in $V_{\rm M}$ at elevated temperatures. $V_{\rm M}$ decreases upon approaching the critical stress value ${\sigma}_{cr}$, reaches the minimum value just above ${\sigma}_{cr}$ and then stays constant. The stress effect on $V_{\rm M}$ is stronger in LBCO-0.135 as compared to LBCO-0.115. We note that the stress effect at 400 mK is largely reduced as compared to 10 K where $V_{\rm M}$ shows strong stress induced suppression. This shows that in the zero-temperature limit, the large-volume-fraction spin-stripe order is compatible with 3D superconducting order. But it seems that the reduction of magnetic fraction and the suppression of the LTT structure is a necessary condition to establish 3D superconducting coherence with optimal $T_{\rm c}$.

\subsection{Discussion}

Our experiments show that when a LBCO sample exhibits a nearly 100 ${\%}$ spin-stripe order $V_{\rm M}$ and a long-range LTT structure, it also exhibits 2D superconductivity and a strongly depressed 3D SC transition. As $V_{\rm M}$ decreases and the LTT structure is suppressed, the 3D SC transition temperature rises. To interpret these results, we first recall that around $x=1/8$ there is a structural phase transition from LTO to an LTT phase. The LTT structure favours the orthogonal stripe order along the $c$ axis. A finite interlayer Josephson coupling would normally be expected to lock the phases of the superconducting wave function between the layers, resulting in 3D SC order. To explain the apparent frustration of such interlayer Josephson coupling, pair-density-wave order within the layers has been proposed \cite{Berg1,Fradkin,Himeda}, which is compatible with both the charge- and spin-stripe orders. Such an anti-phase SC order combined with the LTT structure and the orthogonal stripe ordering, could explain the frustrated Josephson coupling between the layers and the suppression of bulk 3D SC order, while the 2D SC correlations within the CuO$_{2}$ layers still coexist with the static spin-stripe order. The suppression of the tetragonal structure under applied stress implies that stress disfavors an orthogonal stripe arrangement and lifts the geometric frustration. As a result, the ordered fraction $V_{\rm M}$ of spin-stripe order is partly reduced, which seems sufficient to let the system establish 3D superconductivity and to enormously enhance its SC critical temperature. A reduction of the magnetic fraction $V_{\rm M}$ under stress is expected to promote patches of uniform $d$-wave superconductivity. Patches in adjacent layers whose projections overlap, mediate a non-zero interlayer coupling. However, as long as such patches are sparse, the PDW of the stripes dominates the intralayer physics, and the interlayer couplings are thus frustrated. Beyond a critical patch fraction, the superconducting phase is expected to develop uniform long-range order, with a defined phase relationship between the majority of patches. At this point, one expects  $T_{c,3D}$ to coincide with $T_{c,2D}$.

In conclusion, we used muon spin rotation, magnetic susceptibility and X-ray scattering experiments to follow the evolution of the spin-stripe order, superconductivity, and LTO-LTT structural phase transition in LBCO with $x=0.115$ and $x=0.135$ as a function of stress applied within the CuO$_2$ planes (at an angle of 45$^\circ$ to the Cu-O bond direction). Stress induces a full suppression of the LTT structural phase and substantial decrease in the magnetic volume fraction at elevated temperatures, as well as a dramatic rise in the onset of 3D superconductivity on both sides of 1/8 anomaly in the phase diagram. However, the spin-stripe order temperature, as well as the magnetic fraction at 400 mK shows only a modest decrease under stress. Moreover, the onset temperature for the 3D superconductivity and the spin-stripe order are quite similar in the less frustrated large stress regime (beyond the critical stress $\sigma_{cr}$ ${\sim}$ 0.06 GPa), from which we infer that the same kind of electronic interactions are responsible for both phenomena \cite{Tranquadareview}. Our data demonstrate that in-plane strain can be used to affect the phase competition in the striped cuprates. Namely, by strain tuning the magnetic fraction and the crystal structure, one can switch between anti-phase PDW and uniform $d$-wave SC orders. Our results raise fundamental theoretical questions concerning the nature of the strain-stimulated cooperation between the 3D SC and the magnetic order in the stripe phase of cuprates and might shed new light on the high-$T_{\rm c}$ problem.

\textbf{\section{Acknowledgments}}

Muon spin rotation experiments were performed at the Swiss Muon Source S${\mu}$S, Paul Scherrer Institute, Villigen, Switzerland. 
We acknowledge DESY (Hamburg, Germany), a member of the Helmholtz Association HGF, for the provision of experimental facilities. Parts of this research were carried out at PETRA~III and we would like to thank Philipp Glaevecke and Olof Gutowski for assistance in using P21.1. Beamtime was allocated for proposal I-20210503 EC. Work at Brookhaven is supported by the Office of Basic Energy Sciences, Materials Sciences and Engineering Division, U.S. Department of Energy under Contract No.\ DE-SC0012704.\\




\section{METHODS}

\textbf{Sample preparation}: Polycrystalline samples of La$_{2-x}$Ba$_{x}$CuO$_{4}$ with $x$ = 0.115 and $x$ = 0.135 were prepared by the conventional solid-state reaction method using La$_{2}$O$_{3}$, BaCO$_{3}$, and CuO as starting materials. The single-phase character of the samples was checked by powder X-ray diffraction. The single crystals of La$_{2-x}$Ba$_{x}$CuO$_{4}$ with $x$ = 0.115 and $x$ = 0.135 were grown by the traveling solvent floating-zone method \cite{AdachiPRB2001}. All the measurements were performed on samples from the same batch.\\ 
  
\textbf{Uniaxial strain devices}: For the ${\mu}$SR and AC susceptibility measurements, we used the piezoelectric-driven uniaxial pressure apparatus \cite{Hicks,GuguchiaPRLstrain} designed for operation at cryogenic temperatures, where the sample geometry and sample size are suitable for muon spin rotation and relaxation experiments. The apparatus fits into the $Oxford~Heliox$ $^{3}$He cryostat of the general purpose instrument Dolly on the ${\pi}$E1 beamline at the Paul Scherrer Institute. The use of piezoelectric actuators allows for a continuous in situ tunability of the applied pressure. The sample is mounted in a detachable holder, made of titanium that allows rapid sample exchange, as described in detail in our previous work \cite{GuguchiaPRLstrain}. The holder incorporates two pairs of flexures that protect the sample against inadvertent torques and transverse forces. The space around the sample is kept as open as possible, so that muons that miss the sample pass through the holder and are picked up by the veto counter (the purpose of veto counter is to reject muons and their decay positrons that have missed the sample). The sample plates were masked with hematite foils, which strongly depolarizes the incoming muons resulting in loss of asymmetry (signal). We managed to have around ${\sim}$ 40 ${\%}$ of the incoming muons stopped in the sample. The sample holder was attached to the main part of the apparatus, which is called the strain generator, and which contains the piezoelectric actuators.\\

We used a La$_{1.885}$Ba$_{0.115}$CuO$_{4}$ and La$_{1.865}$Ba$_{0.135}$CuO$_{4}$ samples (referred to as LBCO-0.115 and LBCO-0.135 in the following) (7 mm ${\times}$ 4 mm ${\times}$ 0.6 mm) oriented along a crystallographic direction which is off by 45 $^{\circ}$ from the $a$ axis. The sample was fixed on the sample holder by using epoxy (Stycast- 2850 FT) and mounted on a sample holder consisting of grade-2 Ti. A pair of coils (each having ~100 turns) was placed very close to the sample for in situ ac-susceptibility (ACS) measurements \cite{GuguchiaPRLstrain}. These allowed us to determine the $T_{\rm c}$ of LBCO-0.115 and LBCO-0.135 under different stress conditions. The area facing the muon beam was 4 ${\times}$ 4.2 mm$^{2}$.\\ 

The samples were cooled down while keeping the piezoelectric actuators grounded. We applied the compressive stress at 45 K, followed by a system cooling down to 0.4 K. ACS and ${\mu}$SR measurements were carried out upon warming the sample. In order to apply a compressive stress to the sample, a positive voltage was applied in the compression stack ($V_{\rm C}$). To avoid possible electrical discharges of He gas at 45 K, we had to limit $V_{\rm C}$ to +100 V. To achieve a higher compressive force, we kept $V_{\rm C}$ = +100 V and applied a negative voltage in the tension stack ($V_{\rm T}$).\\

For the X-ray experiments, the uniaxial pressure was generated using a dedicated sample stick with a linear actuator generating the force. Integrated feedback mechanisms ensured a constant applied pressure during the temperature scans as described elsewhere \cite{SimutisXray}.
X-ray scattering experiments were carried out at the P21.1 beamline at PETRA-III. The crystal was positioned with the tetragonal [0,0,1] and [1,1,0] directions spanning the scattering plane and the stress was applied at an angle of 45$^\circ$ to the Cu-O bond direction. A single bent-Laue Si monochromator on the (311) reflection was used, resulting in an energy of 101.6 keV. The monochromator crystal was bent towards the sample, increasing the effective incident flux. Scattering from the sample was filtered with a perfect Si analyzer using the (311) reflection in Laue geometry. The resulting scattered light was collected by a PILATUSX CdTe 100k pixel detector, and the intensity was integrated over a region of interest.\\

\subsection{Principles of the ${\mu}$SR technique} Static spin-stripe order in La$_{2-x}$Ba$_{x}$CuO$_{4}$ with $x$ = 0.115 and 
$x$ = 0.135 was studied by means of zero-field (ZF) and weak transverse-field (weak-TF) ${\mu}$SR experiments.
In a ${\mu}$SR experiment an intense beam of 100 ${\%}$ spin-polarized muons is stopped in the sample (see Fig. 1a). The positively charged muons (momentum $p_{\mu}$ = 29 MeV/c) thermalize in the sample at interstitial lattice sites, where they act as magnetic microprobes. In a magnetic material the muon spins precess in the local field $B_{{\rm \mu}}$ at the muon site with a Larmor frequency ${\omega}_{{\rm \mu}}$ = 2${\pi}$ ${\nu}_{{\rm \mu}}$ = $\gamma_{{\rm \mu}}$$B_{{\rm \mu}}$ [muon gyromagnetic ratio $\gamma_{{\rm \mu}}$/(2${\pi}$) = 135.5 MHz T$^{-1}$]. In a ZF ${\mu}$SR experiment, positive muons implanted into a sample serve as an extremely sensitive local probe to detect small internal magnetic fields and ordered magnetic volume fractions in the bulk of magnetic materials. Thus, ${\mu}$SR is a particularly powerful tool to study inhomogeneous magnetism in materials. 

The positive muons $\mu^{+}$ implanted into the sample decay after a mean life time of ${\tau}$$_{\mu}$ = 2.2 ${\mu}$s, emitting a fast positron $e^{+}$, preferentially along the muon spin direction. Various detectors placed around the sample track the incoming $\mu^{+}$ and the outgoing $e^{+}$. When the muon detector records the arrival of a $\mu^{+}$ in the specimen, the electronic clock starts. The clock stops when the decay positron $e^{+}$ is registered in one of the $e^{+}$ detectors, and the measured time interval is stored in a histogramming memory. In this way a positron-count versus time histogram is formed. A muon decay event requires that within a certain time interval after a $\mu^{+}$ arrival an $e^{+}$ is detected. This time interval extends usually over several muon lifetimes (e.g., 10\,$\mu$s). In the Dolly instrument the sample is surrounded by four positron detectors (with respect to the muon-beam direction): Forward, Backward, Left, and Right. After several millions of muons stopped in the sample, one at a time, one obtains a histogram for the positrons $e^{+}$, revealed in the forward ($N_{\rm F}$), the backward ($N_{\rm B}$), the left ($N_{\rm L}$), and the right ($N_{\rm R}$) detectors. Ideally, the histogram counts are described by:

\begin{equation} 
N_{{\alpha}}(t)=N_{0}e^{-\frac{t}{\tau_{\mu}}}[1+A_{0}\vec{P}(t)\vec{n}_{\alpha}]+N_{\mathrm{bg}}. 
\end{equation}
  
 Here, the exponential factor accounts for the radioactive muon decay.
$\vec{P}$($t$) is the muon-spin polarization function with the unit vector 
${\vec{n}_{\alpha}}$ (${\alpha}$ = F, B, L, R) along the direction of the positron detector. $N_{\rm 0}$ is a parameter proportional with the number of the recorded events. $N_{\mathrm{bg}}$ is a background contribution due to uncorrelated starts and stops. $A_{0}$ is the initial decay asymmetry, which depends on different experimental factors, such as the detector solid angle, absorption, and scattering of positrons in the material. Typical values of $A_{0}$ are between 0.2 and 0.3. 

  Since the positrons are emitted predominantly in the direction of the muon spin (here precessing with ${\omega_{\mu}}$), the forward and backward detectors will detect a signal oscillating with the same frequency. In order to remove the exponential decay (which reflects simply the muon's finite lifetime), the so-called reduced asymmetry signal $A$(t) is calculated:
\begin{equation} 
A(t)=\frac{N_{F,L}(t)-N_{B,R}(t)}{N_{F,L}(t)+N_{B,R}(t)}=A_{0}P(t),
\end{equation}
where, $N_{F,L}$(t) and $N_{B,R}$(t) are the number of positrons detected in the Forward(Left) and Backward(Right) detectors, respectively. The quantities $A(t)$ and $P(t)$ depend sensitively on the
static spatial distribution and the fluctuations of the magnetic environment of the muons.
As such, these functions reflect the physics of the investigated system \cite{Dalmas1997}.\\

\subsection{Analysis of ZF ${\mu}$SR data} The ${\mu}$SR signals (Figure 3b) over the whole temperature range were analyzed by decomposing the asymmetry signal into a magnetic and a nonmagnetic contribution: 
\begin{equation}
\begin{split}
P_{ZF}(t)=V_{m}\Bigg[{f_{\alpha} e^{-\lambda_{T}t}J_0(\gamma_{\mu}B_{int}t)}+(1-f_{\alpha})e^{-\lambda_{L}t}\Bigg]  \\
  +(1-V_{m})e^{-\lambda_{nm}t}.
\label{eq1}
\end{split}
\end{equation}
Here, $V_{\rm M}$ denotes the relative volume of the magnetic fraction, and $B_{\rm int}$ is the maximal value of the internal field distribution (Overhauser distribution). ${\lambda}_{\rm T}$ and ${\lambda}_{\rm L}$ are the depolarization rates representing the transverse and the longitudinal relaxing components of the magnetic parts of the sample. $f$ and $(1-f)$ are the fractions of the oscillating and non-oscillating components of the magnetic ${\mu}$SR signal. $J_{0}$ is the zeroth-order Bessel function of the first kind. This is characteristic of an incommensurate spin density wave, as well as of broad internal field distributions with fields ranging from zero to a maximal field and has been regularly observed in cuprates with static spin-stripe order \cite{GuguchiaNPJ,Nachumi}.
${\lambda_{nm}}$ is the relaxation rate of the nonmagnetic part of the sample, where the spin-stripe order is absent. All the ${\mu}$SR time spectra (both ZF and TF) were analyzed using the free software package {\tt musrfit} \cite{AndreasSuter}.\\

\subsection{Analysis of weak-TF ${\mu}$SR data} The weak-TF asymmetry spectra, shown in Fig. 3a, were analyzed by using the function:

\begin{equation}
\begin{aligned}P_{TF}(t)= P_{TF}(0)\exp(-\lambda t)\cos(\omega t + \phi),
\label{eq1}
\end{aligned}
\end{equation}
where $t$ is the time after muon implantation, $P_{\rm TF}$($t$) is the time-dependent polarization, $P_{\rm TF}$(0) is the initial polarization (amplitude) of the low frequency oscillating component (related to the paramagnetic volume fraction), ${\lambda}$ is an exponential damping rate due to paramagnetic spin fluctuations and/or nuclear dipolar moments, ${\omega}$ is the Larmor precession frequency, which is proportional to the strength of the external transverse magnetic field, and ${\phi}$ is a phase offset. As it is standard for the analysis of weak-TF data from magnetic samples the zero for $P_{\rm TF}$($t$) was allowed to vary for each temperature to deal with the asymmetry baseline shift occurring in magnetically ordered samples. From these refinements, the magnetically ordered volume fraction at each temperature $T$ was determined by $V_{\rm M}$ = 1  -- $P_{\rm TF}$(0)($T$). In the paramagnetic phase at high temperature $P_{\rm TF}$(0)($T$ ${\textgreater}$ $T_{\rm so}$) = 1.\\ 

In general, weak-TF signal consists of long-lived oscillations with an applied field and strongly damped oscillations from muons in magnetically ordered regions experiencing a broad field distribution due to the vector sum of the applied and internal fields. It is clear from Fig. 3b that in ZF-${\mu}$SR, the depolarisation occurs on a ${\sim}$ 0.5 ${\mu}$s time scale. However, in TF-${\mu}$SR data, due to a distribution in the angle between the applied and internal fields, the inhomegeneity of the field magnitude is large in the magnetically ordered regions, which causes a strong dephasing of the muon spins. Because of this and due to a strong data binning (which averages the signal in the time bins), the strongly damped signal from muons in the magnetic patches of the sample is not visible in the weak-TF ${\mu}$SR spectra.\\


\end{document}